# 4, 8, 32, 64 bit Substitution Box generation using Irreducible or Reducible Polynomials over Galois Field GF($p^q$) for Smart Applications.

Sankhanil Dey[1] and Ranjan Ghosh[2],
sdrpe_rs@caluniv.ac.in[1], rghosh47@yahoo.co.in[2],
Institute of Radio Physics and Electronics[1,2],
University of Calcutta.

**Abstract.** Substitution Box or S-Box had been generated using 4-bit Boolean Functions (BFs) for Encryption and Decryption Algorithm of Lucifer and Data Encryption Standard (DES) in late sixties and late seventies respectively. The S-Box of Advance Encryption Standard have also been generated using Irreducible Polynomials over Galois field GF($2^8$) adding an additive constant in early twenty first century. In this paper Substitution Boxes have been generated from Irreducible or Reducible Polynomials over Galois field GF($p^q$). Binary Galois fields have been used to generate Substitution Boxes. Since the Galois Field Number or the Number generated from coefficients of a polynomial over a particular Binary Galois field ($2^q$) is similar to $\log_2{}^{q+1}$ bit BFs. So generation of $\log_2{}^{q+1}$ bit S-Boxes is Possible. Now if p = prime or non-prime number then generation of S-Boxes is possible using Galois field GF ($p^q$). where, q = p-1.

1. **Introduction.** Polynomials over Finite field or Galois field GF($p^q$) have been of utmost importance in Public Key Cryptography [1]. The polynomials over Finite field or Galois field GF($p^q$) that cannot be factored into polynomials with less de-gree of d and q-d where d ={1,2,…..,(q-1)/2} have been termed as Irreducible polynomials over Finite field or Galois field $GF(p^q)$ and the rest have been termed as Reducible polynomials over Finite field or Galois field $GF(p^q)$ [2]. The polynomials over Galois field $GF(p^q)$ with coefficient of the highest degree term as 1 have been termed as Monic polynomials Galois field $GF(p^q)$and rest have been termed as Non-monic Polynomials Galois field $GF(p^q)$ [3]. The polynomials Galois field $GF(p^q)$ with degree q have been termed as Basic Polynomials or BPs over Galois field $GF(p^q)$ and Polynomials with degree q-1 have been termed as Elemental Polynomials or EPs over Galois field $GF(p^q)$ [4].

q bit proper Substitution box or S-Box have $2^q$ elements in an array where each element is unique and distinct and arranged in a random fashion varies from 0 to q. Polynomials over Galois field $GF(p^q)$ have been termed as binary polynomials if p = 2. The binary number constructed with q = 0 at LSB and q = q at MSB has been termed as binary Coefficient Number or BCN of $\log_2{}^{q+1}$ bits. The Binary Coefficient Number or BCN over Galois field $GF(p^q)$ has been similar with $\log_2{}^{q+1}$ bit BFs. The $\log_2{}^{q+1}$ bit S-Boxes have been generated using $\log_2{}^{q+1}$ bit BCNs. In this paper proper 4, 5, 6, 7, and 8 bit S-Boxes have been generated using BCNs and the procure has been continued as a future scope to generate 16 and 32 bit S-Boxes. The non-repeated Coefficients of BPs over Galois field $GF(p^q)$, where $P = 2^{(\log_2{}^{q+1})}$ and q = p-1 have been used to generate $\log_2{}^{q+1}$ bit S-Boxes. In this paper proper 4, 5, 6, 7, and 8 bit S-Boxes have been generated using BCNs and the procure has been continued as a future scope to generate 16 and 32 bit S-Boxes.

Main goal of the Smart Object For Intelligent Application (SOFIA) project is to make "information" in the physical world available for smart services - connecting physical world with information world [5][6][7]. Full access to information present in the embedded computing devices has a potential for large impact on the daily lives of people living in this environment[8][9][10]. Sharing Information safely has been of utmost importance in SOFIA. Modern block ciphers have been of utmost importance in doing so. The substitution boxes are the major part of ancient as well as modern block ciphers[11][12].

In this paper Polynomials over Galois field GF($p^q$) and Substitution Boxes have been reviewed in section.2. The generation of 4 and 8 bit S-Boxes using BCNs have been elaborated in section 3. The generation of S-Boxes of 4 and 8 bit using Coefficients of Non-binary Galois Field Polynomials have been depicted in section 4. Conclusion, Acknowledgement and Reference has been given in section 5, 6 and 7.

**1. Polynomials over Galois field GF($p^q$) and $\log_2{}^{q+1}$ bit S-Boxes.** In this section the sub section 2.1. has been devoted to a small review of Polynomials. The sub section 2.2. has been of Utmost importance since in it A four bit bijective Crypto or Proper S-Box has been defined in brief. At last in sub section 2.3. The equation among $2^{15}$ Galois field Polynomials and a 4-bit Bijective Crypto S-Box has been elaborated in details.

**2.1. Polynomials over Galois field GF($p^q$)**. Polynomials over Galois field GF($p^q$) have been of utmost importance in Cryptographic Applications. Polynomials with degree q have been termed as Basic Polynomials and Polynomials with degree less than q have been termed as Elemental Polynomials over Galois field GF($p^q$). Polynomials with leading coefficient as 1 have been termed as Monic Polynomials irrespective of BPs and EPs. An example, of the said criteria have been described as follows, the Example of Basic Polynomial or BP over Galois field GF($p^q$) has been given below,

$BP(x) = co_q x^q + co_{q-1} x^{q-1} + co_{q-1} x^{q-2} + \ldots\ldots\ldots\ldots\ldots\ldots\ldots\ldots + co_2 x^2 + co_1 x^1 + a_0 \ldots\ldots\ldots\ldots\ldots\ldots\ldots\ldots\ldots\ldots\ldots$(i)

In equation (i) BP(x) has been represented as Basic Polynomial over Galois field GF($p^q$)since the highest degree term of the said Polynomial over Galois field GF($p^q$) is €q. The BP has been called as a Monic BP if $co_{q\,=}$ 1. The number of Terms in a BP over Galois field GF($p^q$) has been 0 to q i.e. (q+1). The number of possible values of a particular coefficient $co_{q,}$ where 0≤q≤q has been from 0 to p i.e. € (p+1). If the value of q has been <q then The Polynomial over Galois field GF($p^q$) has been termed as Elemental Polynomial over Galois field GF($p^q$). If a BP over Galois field GF($p^q$) can be factored into two non-constant EPs then the BP can be termed as Reducible Polynomials over Galois field GF($p^q$). If the two factor of a BP over Galois field GF($p^q$) have been the BP itself and a constant Polynomial then The BP have been said as an Irreducible Polynomial over Galois field GF($p^q$).

**1.2. 4-bit Crypto S-Boxes:** A 4-bit bijective Crypto S-Box can be written as Follows, where the each element of the first row of Table.1, entitled as index, are the position of each element of the S-Box within the given S-Box and the elements of the 2nd row, entitled as S-Box, are the elements of the given Substitution Box. It can be concluded that the 1st row is fixed for all possible bijective crypto S-Boxes. The values of each element of the 1st row are distinct, unique and vary between 0 and F. The values of the each element of the 2nd row of a bijective crypto S-Box are also distinct and unique and also vary between 0 and F. The values of the elements of the fixed 1st row are sequential and monotonically increasing where for the 2nd row they can be sequential or partly sequential or non- sequential. Here the given Substitution Box is the 1st 4-bit S-Box of the 1st S-Box out of 8 of Data Encryption Standard [13][14][15].

| **Row** | **Col** | 1 | 2 | 3 | 4 | 5 | 6 | 7 | 8 | 9 | A | B | C | D | E | F | G |
|---|---|---|---|---|---|---|---|---|---|---|---|---|---|---|---|---|---|
| 1 | **Index** | 0 | 1 | 2 | 3 | 4 | 5 | 6 | 7 | 8 | 9 | A | B | C | D | E | F |
| 2 | **S-Box** | E | 4 | D | 1 | 2 | F | B | 8 | 3 | A | 6 | C | 5 | 9 | 0 | 7 |

**Table.1. 4-bit bijective Crypto S-Box.**

**1.3. Relation between 4-bit S-Boxes and Polynomials over Galois field GF ($2^{15}$).** Index of Each element of a 4-bit bijective crypto S-Box and the element itself is a hexadecimal number and that can be converted into a 4-bit bit sequence. From row 2 through 5 and row 7 through A of each column from 1 through G of Table.2. shows the 4-bit bit sequences of the corresponding hexadecimal numbers of the index of each element of the given S-Box and each element of the S-Box itself. Each row from 2 through 5 and 7 through A from column 1 through G constitutes a 16 bit, bit sequence that is a Basic Polynomial over Galois field GF($2^{15}$). column 1 through G of Row 2 is termed as 4th IGFP, Row 3 is termed as 3rd IGFP Row 4 is termed as 2nd IGFP and Row 5 is termed as IGFP whereas column 1 through G of Row 7 is termed as 4th OGFP, Row 8 is termed as 3rd OGFP, Row 9 is termed as 2nd OGFP and Row A is termed as 1st OGFP. The decimal equivalent of each IGFP and OGFP are noted at column H of respective rows. Where IGFP stands for Input Galois Field Polynomial and OGFP stands for Output Galois Field Polynomials. The respective Polynomials have been shown in Row 1 through 8 of column 3 of Table.3.

| **2.** | **Column** | 1 | 2 | 3 | 4 | 5 | 6 | 7 | 8 | 9 | A | B | C | D | E | F | G | **H.** Decimal |
|---|---|---|---|---|---|---|---|---|---|---|---|---|---|---|---|---|---|---|
| 1 | **Index** | **0** | **1** | **2** | **3** | **4** | **5** | **6** | **7** | **8** | **9** | **A** | **B** | **C** | **D** | **E** | **F** | |
| 2 | **IBF4** | 0 | 0 | 0 | 0 | 0 | 0 | 0 | 0 | 1 | 1 | 1 | 1 | 1 | 1 | 1 | 1 | 00255 |
| 3 | **IBF3** | 0 | 0 | 0 | 0 | 1 | 1 | 1 | 1 | 0 | 0 | 0 | 0 | 1 | 1 | 1 | 1 | 03855 |
| 4 | **IBF2** | 0 | 0 | 1 | 1 | 0 | 0 | 1 | 1 | 0 | 0 | 1 | 1 | 0 | 0 | 1 | 1 | 13107 |
| 5 | **IBF1** | 0 | 1 | 0 | 1 | 0 | 1 | 0 | 1 | 0 | 1 | 0 | 1 | 0 | 1 | 0 | 1 | 21845 |
| 6 | **S-Box** | E | 4 | D | 1 | 2 | F | B | 8 | 3 | A | 6 | C | 5 | 9 | 0 | 7 | |
| 7 | **OBF4** | 1 | 0 | 1 | 0 | 0 | 1 | 1 | 1 | 0 | 1 | 0 | 1 | 0 | 1 | 0 | 0 | 42836 |
| 8 | **OBF3** | 1 | 1 | 1 | 0 | 0 | 1 | 0 | 0 | 0 | 0 | 1 | 1 | 1 | 0 | 0 | 1 | 58425 |
| 9 | **OBF2** | 1 | 0 | 0 | 0 | 1 | 1 | 1 | 0 | 1 | 1 | 1 | 0 | 0 | 0 | 0 | 1 | 36577 |
| A | **OBF1** | 0 | 0 | 1 | 1 | 0 | 1 | 1 | 0 | 1 | 0 | 0 | 0 | 1 | 1 | 0 | 1 | 13965 |

**Table.2. Input and Output BCNs of the Substitution Box**

| **Col** | 1 | 2 | 3 |
|---|---|---|---|
| | **Index** | **DCM Eqv.** | **Polynomials over Galois Field GF($2^{15}$).** |
| 1 | IGFP4 | 00255 | $BP(x) = x^7+x^6+x^5+x^4+x^3+x^2+x^1+1$. |
| 2 | IGFP3 | 03855 | $BP(x) = x^{11}+x^{10}+x^9+x^8+x^3+x^2+x^1+1$. |
| 3 | IGFP2 | 13107 | $BP(x) = x^{13}+x^{12}+x^9+x^8+x^5+x^4+x^1+1$. |
| 4 | IGFP1 | 21845 | $BP(x) = x^{14}+x^{12}+x^{10}+x^8+x^6+x^4+x^2+1$. |
| 5 | OGFP4 | 42836 | $BP(x) = x^{15}+x^{13}+x^{10}+x^9+x^8+x^6+x^4+ x^2$. |
| 6 | OGFP3 | 58425 | $BP(x) = x^{15}+x^{14}+x^{13}+x^{10}+x^5+x^4+x^3+1$. |
| 7 | OGFP2 | 36577 | $BP(x) = x^{15}+x^{11}+x^{10}+x^9+x^7+x^6+x^5+1$. |
| 8 | OGFP1 | 13965 | $BP(x) = x^{13}+x^{12}+x^{10}+x^9+x^7+x^3+x^2+1$. |

**Table.3. Respective Polynomials of IGFP4 through IGFP1 and OGFP4 through OGFP1**

## 3. 4 and 8 bit S-Box Generation by respective BCNs over Binary Galois Field GF($2^q$) where q €15 and 255 respectively.

In this paper 4 and 8 bit Identity S-Boxes have been taken for example for generation of 4 and 8 bit S-Boxes over Binary Galois Fields GF($2^q$) where q €15 and 255 respectively. The generation of Identity 4-bit S-Box from four BCNs over Binary Galois Field GF($2^{15}$) have been elaborated in sub section 3.1 and The generation of Identity 8-bit S-Box from Eight BCNs over Binary Galois Field GF($2^{255}$) have been elaborated in sub section 3.2. The Algorithm for generation of $\log_2{}^{q+1}$ bit S-Boxes over Binary Galois Field GF($2^q$) has been depicted with Time Complexity of the algorithm in sub section 3.3.

### 3.1. Generation of 4-bit Identity Crypto S-Box from four Polynomials over Binary Galois Field GF($2^{15}$).

The Concerned 4-bit Identity S-Box has been shown in table.4 where each element of the first row of Table.4, entitled as index, are the position of each element of the S-Box within the given S-Box and the elements of the 2nd row, entitled as S-Box, are the elements of the given Identity Substitution Box. It can be concluded that the 1st row is fixed for all possible bijective crypto S-Boxes. The values of each element of the 1st row are distinct, unique and vary between 0 and F. The values of the each element of the 2nd row of the Identity crypto S-Box are also distinct and unique and also vary between 0 and F. The values of the elements of the fixed 1st row are sequential and monotonically increasing where for the 2nd row, they can be sequential or partly sequential or non- sequential. Here the given Substitution Box is the 4-bit Identity Crypto S-Box.

| **Row** | **Column** | 1 | 2 | 3 | 4 | 5 | 6 | 7 | 8 | 9 | A | B | C | D | E | F | G |
|---|---|---|---|---|---|---|---|---|---|---|---|---|---|---|---|---|---|
| 1 | **Index** | 0 | 1 | 2 | 3 | 4 | 5 | 6 | 7 | 8 | 9 | A | B | C | D | E | F |
| 2 | **S-Box** | 0 | 1 | 2 | 3 | 4 | 5 | 6 | 7 | 8 | 9 | A | B | C | D | E | F |

**Table.4. 4-bit Identity Crypto S-Box.**

Index of Each element of a 4-bit bijective crypto S-Box and the element itself is a hexadecimal number and that can be converted into a 4-bit bit sequence. From row 2 through 5 and row 7 through A of each column from 1 through G of Table.5. shows the 4-bit bit sequences of the corresponding hexadecimal numbers of the index of each element of the given S-Box and each element of the S-Box itself. Each row from 2 through 5 and 7 through A from column 1 through G constitutes a 16 bit, bit sequence that is a Basic Polynomial over Galois field GF($2^{15}$). column 1 through G of Row 2 is termed as 4th IGFP, Row 3 is termed as 3rd IGFP Row 4 is termed as 2nd IGFP and Row 5 is termed as IGFP whereas column 1 through G of Row 7 is termed as 4th OGFP, Row 8 is termed as 3rd OGFP, Row 9 is termed as 2nd OGFP and Row A is termed as 1st OGFP. The decimal equivalent of each IGFP and OGFP are noted at column H of respective rows. Where IGFP stands for Input Galois Field Polynomial and OGFP stands for Output Galois Field Polynomials. The respective Polynomials have been shown in Row 1 through 8 of column 3 of Table.6.

| **Row** | **Column** | 1 | 2 | 3 | 4 | 5 | 6 | 7 | 8 | 9 | A | B | C | D | E | F | G | **H.** Decimal |
|---|---|---|---|---|---|---|---|---|---|---|---|---|---|---|---|---|---|---|
| 1 | **Index** | **0** | **1** | **2** | **3** | **4** | **5** | **6** | **7** | **8** | **9** | **A** | **B** | **C** | **D** | **E** | **F** | |
| 2 | **IBCN4** | 0 | 0 | 0 | 0 | 0 | 0 | 0 | 0 | 1 | 1 | 1 | 1 | 1 | 1 | 1 | 1 | 00255 |
| 3 | **IBCN3** | 0 | 0 | 0 | 0 | 1 | 1 | 1 | 1 | 0 | 0 | 0 | 0 | 1 | 1 | 1 | 1 | 03855 |
| 4 | **IBCN2** | 0 | 0 | 1 | 1 | 0 | 0 | 1 | 1 | 0 | 0 | 1 | 1 | 0 | 0 | 1 | 1 | 13107 |
| 5 | **IBCN1** | 0 | 1 | 0 | 1 | 0 | 1 | 0 | 1 | 0 | 1 | 0 | 1 | 0 | 1 | 0 | 1 | 21845 |
| 6 | **S-Box** | **0** | **1** | **2** | **3** | **4** | **5** | **6** | **7** | **8** | **9** | **A** | **B** | **C** | **D** | **E** | **F** | |
| 7 | **OBCN4** | 0 | 0 | 0 | 0 | 0 | 0 | 0 | 0 | 1 | 1 | 1 | 1 | 1 | 1 | 1 | 1 | 00255 |
| 8 | **OBCN3** | 0 | 0 | 0 | 0 | 1 | 1 | 1 | 1 | 0 | 0 | 0 | 0 | 1 | 1 | 1 | 1 | 03855 |
| 9 | **OBCN2** | 0 | 0 | 1 | 1 | 0 | 0 | 1 | 1 | 0 | 0 | 1 | 1 | 0 | 0 | 1 | 1 | 13107 |
| A | **OBCN1** | 0 | 1 | 0 | 1 | 0 | 1 | 0 | 1 | 0 | 1 | 0 | 1 | 0 | 1 | 0 | 1 | 21845 |

**Table.5. Input and Output BCNs of the Identity Substitution Box**

| **Col** | 1 | 2 | 3 |
|---|---|---|---|
| | **Index** | **DCM Eqv.** | **Polynomials over Galois Field GF($2^{15}$).** |
| 1 | IGFP4 | 00255 | BP(x) = $x^7+x^6+x^5+x^4+x^3+x^2+x^1+1$. |
| 2 | IGFP3 | 03855 | BP(x) = $x^{11}+x^{10}+x^9+x^8+x^3+x^2+x^1+1$. |
| 3 | IGFP2 | 13107 | BP(x) = $x^{13}+x^{12}+x^9+x^8+x^5+x^4+x^1+1$. |
| 4 | IGFP1 | 21845 | BP(x) = $x^{14}+x^{12}+x^{10}+x^8+x^6+x^4+x^2+1$. |
| 5 | OGFP4 | 00255 | BP(x) = $x^7+x^6+x^5+x^4+x^3+x^2+x^1+1$. |
| 6 | OGFP3 | 03855 | BP(x) = $x^{11}+x^{10}+x^9+x^8+x^3+x^2+x^1+1$. |
| 7 | OGFP2 | 13107 | BP(x) = $x^{13}+x^{12}+x^9+x^8+x^5+x^4+x^1+1$. |
| 8 | OGFP1 | 21845 | BP(x) = $x^{14}+x^{12}+x^{10}+x^8+x^6+x^4+x^2+1$. |

**Table.6. Respective Polynomials of IGFP4 through IGFP1 and OGFP4 through OGFP1.**

### 3.2. Generation of 8-bit Identity Crypto S-Box from Eight Polynomials over Binary Galois Field GF($2^{255}$).

The Concerned 8-bit Identity S-Box has been shown in table.7 where each element of the first row of Table.7, entitled as index, are the position of each element of the S-Box within the given S-Box and the elements of the column 1 through G of 2nd to 17th row, entitled as S-Box, are the elements of the given 8-bit Identity Substitution Box sequentially. It can be concluded that the 1st row is fixed for all possible 8-bit bijective crypto S-Boxes. The values of each element of the 1st row are distinct, unique and vary between 0 and F. The values of the each element of the column 1 through G of 2nd row to 17th row of the 8-bit Identity crypto S-Box are also distinct and unique and vary between 0 and 256. The values of the elements of the fixed 1st row are sequential and monotonically increasing where for the 2nd to 17th row, they can be sequential or partly sequential or non- sequential. Here the given Substitution Box is the 8-bit Identity Crypto S-Box.

| **Row** | **Column** | 1 | 2 | 3 | 4 | 5 | 6 | 7 | 8 | 9 | A | B | C | D | E | F | G |
|---|---|---|---|---|---|---|---|---|---|---|---|---|---|---|---|---|---|
| 1 | **Index** | 0 | 1 | 2 | 3 | 4 | 5 | 6 | 7 | 8 | 9 | A | B | C | D | E | F |
| 2 | **S-Box** | 0 | 1 | 2 | 3 | 4 | 5 | 6 | 7 | 8 | 9 | 10 | 11 | 12 | 13 | 14 | 15 |
| 3 | | 16 | 17 | 18 | 19 | 20 | 21 | 22 | 23 | 24 | 25 | 26 | 27 | 28 | 29 | 30 | 31 |
| 4 | | 32 | 33 | 34 | 35 | 36 | 37 | 38 | 39 | 40 | 41 | 42 | 43 | 44 | 45 | 46 | 47 |
| 5 | | 48 | 49 | 50 | 51 | 52 | 53 | 54 | 55 | 56 | 57 | 58 | 59 | 60 | 61 | 62 | 63. |
| 6 | | 64 | 65 | 66 | 67 | 68 | 69 | 70 | 71 | 72 | 73 | 74 | 75 | 76 | 77 | 78 | 79 |
| 7 | | 80 | 81 | 82 | 83 | 84 | 85 | 86 | 87 | 88 | 89 | 90 | 91 | 92 | 93 | 94 | 95 |
| 8 | | 96 | 97 | 98 | 99 | 100 | 101 | 102 | 103 | 104 | 105 | 106 | 107 | 108 | 109 | 110 | 111 |
| 9 | | 112 | 113 | 114 | 115 | 116 | 117 | 118 | 119 | 120 | 121 | 122 | 123 | 124 | 125 | 126 | 127 |
| 10 | | 128 | 129 | 130 | 131 | 132 | 133 | 134 | 135 | 136 | 137 | 138 | 139 | 140 | 141 | 142 | 143 |
| 11 | | 144 | 145 | 146 | 147 | 148 | 149 | 150 | 151 | 152 | 153 | 154 | 155 | 156 | 157 | 158 | 159 |
| 12 | | 160 | 161 | 162 | 163 | 164 | 165 | 166 | 167 | 168 | 169 | 170 | 171 | 172 | 173 | 174 | 175 |
| 13 | | 176 | 177 | 178 | 179 | 180 | 181 | 182 | 183 | 184 | 185 | 186 | 187 | 188 | 189 | 190 | 191 |
| 14 | | 192 | 193 | 194 | 195 | 196 | 197 | 198 | 199 | 200 | 201 | 202 | 203 | 204 | 205 | 206 | 207 |
| 15 | | 208 | 209 | 210 | 211 | 212 | 213 | 214 | 215 | 216 | 217 | 218 | 219 | 220 | 221 | 222 | 223 |
| **Row** | | 1 | 2 | 3 | 4 | 5 | 6 | 7 | 8 | 9 | A | B | C | D | E | F | G |
| 16 | | 224 | 225 | 226 | 227 | 228 | 229 | 230 | 231 | 232 | 233 | 234 | 235 | 236 | 237 | 238 | 239 |
| 17 | | 240 | 241 | 242 | 243 | 244 | 245 | 246 | 247 | 248 | 249 | 250 | 251 | 252 | 253 | 254 | 255 |

**Table.7. 8-bit Identity Crypto S-Box.**

Index of Each element of an 8-bit bijective crypto S-Box and the element itself is a hexadecimal number and that can be converted into a 256-bit long 8 bit bit sequence. From row 2 through 9 and row A through H of column 2 of Table.8. shows the 8-bit bit sequences of the corresponding hexadecimal numbers of the index of each element of the given S-Box and each element of the S-Box itself. Each row from 2 through 9 and A through H of column 2 constitutes a 256 bit, bit sequence that is a Basic Polynomial over Galois field GF($2^{255}$). column 2 of Row 2 is termed as 8$^{th}$ IGFP, Row 3 is termed as 7$^{th}$ IGFP, Row 4 is termed as 6$^{th}$ IGFP ,Row 5 is termed as 5$^{th}$ IGFP, Row 6 is termed as 4$^{th}$ IGFP, Row 7 is termed as 3$^{rd}$ IGFP, Row 8 is termed as 2$^{nd}$ IGFP and Row 9 is termed as 1$^{st}$ IGFP whereas column 2 of Row A is termed as 8$^{th}$ OGFP, Row B is termed as 7$^{th}$ OGFP, Row C is termed as 6$^{th}$ OGFP, Row D is termed as 5$^{th}$ OGFP, Row E is termed as 4$^{th}$ OGFP, Row F is termed as 3$^{rd}$ OGFP, Row G is termed as 2$^{nd}$ OGFP and Row H is termed as 1st IGFP. The Binary Coefficient Number of each IGFP and OGFP from MSB[256th bit] to LSB[0th bit] have been given in corresponding rows of each IGFP and OGFP. Where IGFP stands for Input Galois Field Polynomial and OGFP for Output Galois Field Polynomials. The respective Polynomial for IGFP8 and OGFP8 has been shown in Table.9.

| Row | Col. | MSB Polynomials (BCNs)[col.2] LSB |
|---|---|---|
| 1 | **1** | |
| 2 | **IGFP8** | 0000000000000000000000000000000000000000000000000000000000000000<br>0000000000000000000000000000000000000000000000000000000000000000<br>1111111111111111111111111111111111111111111111111111111111111111<br>1111111111111111111111111111111111111111111111111111111111111111 |
| 3 | **IGFP 7** | 0000000000000000000000000000000000000000000000000000000000000000<br>1111111111111111111111111111111111111111111111111111111111111111<br>0000000000000000000000000000000000000000000000000000000000000000<br>1111111111111111111111111111111111111111111111111111111111111111 |
| 4 | **IGFP 6** | 0000000000000000000000000000000011111111111111111111111111111111<br>0000000000000000000000000000000011111111111111111111111111111111<br>0000000000000000000000000000000011111111111111111111111111111111<br>0000000000000000000000000000000011111111111111111111111111111111 |
| 5 | **IGFP 5** | 0000000000000000111111111111111100000000000000001111111111111111<br>0000000000000000111111111111111100000000000000001111111111111111<br>0000000000000000111111111111111100000000000000001111111111111111<br>0000000000000000111111111111111100000000000000001111111111111111 |
| 6 | **IGFP 4** | 0000000011111111000000001111111100000000111111110000000011111111<br>0000000011111111000000001111111100000000111111110000000011111111<br>0000000011111111000000001111111100000000111111110000000011111111<br>0000000011111111000000001111111100000000111111110000000011111111 |
| 7 | **IGFP 3** | 0000111100001111000011110000111100001111000011110000111100001111<br>0000111100001111000011110000111100001111000011110000111100001111<br>0000111100001111000011110000111100001111000011110000111100001111<br>0000111100001111000011110000111100001111000011110000111100001111 |
| 8 | **IGFP 2** | 0011001100110011001100110011001100110011001100110011001100110011<br>0011001100110011001100110011001100110011001100110011001100110011<br>0011001100110011001100110011001100110011001100110011001100110011<br>0011001100110011001100110011001100110011001100110011001100110011 |
| 9 | **IGFP 1** | 0101010101010101010101010101010101010101010101010101010101010101<br>0101010101010101010101010101010101010101010101010101010101010101<br>0101010101010101010101010101010101010101010101010101010101010101<br>0101010101010101010101010101010101010101010101010101010101010101 |
| A | **OGFP8** | 0000000000000000000000000000000000000000000000000000000000000000<br>0000000000000000000000000000000000000000000000000000000000000000<br>1111111111111111111111111111111111111111111111111111111111111111<br>1111111111111111111111111111111111111111111111111111111111111111 |
| B | **OGFP 7** | 0000000000000000000000000000000000000000000000000000000000000000<br>1111111111111111111111111111111111111111111111111111111111111111<br>0000000000000000000000000000000000000000000000000000000000000000<br>1111111111111111111111111111111111111111111111111111111111111111 |
| C | **OGFP 6** | 0000000000000000000000000000000011111111111111111111111111111111<br>0000000000000000000000000000000011111111111111111111111111111111<br>0000000000000000000000000000000011111111111111111111111111111111<br>0000000000000000000000000000000011111111111111111111111111111111 |
| D | **OGFP 5** | 0000000000000000111111111111111100000000000000001111111111111111<br>0000000000000000111111111111111100000000000000001111111111111111<br>0000000000000000111111111111111100000000000000001111111111111111<br>0000000000000000111111111111111100000000000000001111111111111111 |
| E | **OGFP 4** | 0000000011111111000000001111111100000000111111110000000011111111<br>0000000011111111000000001111111100000000111111110000000011111111<br>0000000011111111000000001111111100000000111111110000000011111111<br>0000000011111111000000001111111100000000111111110000000011111111 |
| F | **OGFP 3** | 0000111100001111000011110000111100001111000011110000111100001111<br>0000111100001111000011110000111100001111000011110000111100001111<br>0000111100001111000011110000111100001111000011110000111100001111<br>0000111100001111000011110000111100001111000011110000111100001111 |
| G | **OGFP 2** | 0011001100110011001100110011001100110011001100110011001100110011<br>0011001100110011001100110011001100110011001100110011001100110011<br>0011001100110011001100110011001100110011001100110011001100110011<br>0011001100110011001100110011001100110011001100110011001100110011 |
| H | **OGFP 1** | 0101010101010101010101010101010101010101010101010101010101010101<br>0101010101010101010101010101010101010101010101010101010101010101<br>0101010101010101010101010101010101010101010101010101010101010101<br>0101010101010101010101010101010101010101010101010101010101010101 |

**Table.8. BCNs for 8 IGFPs and OGFPs.**

| BCNs of | Polynomial |
|---|---|
| IGFP8 &OGFP8 | $x^{127}+ x^{126}+ x^{125}+ x^{124}+ x^{123}+ x^{122}+ x^{121}+ x^{120}+ x^{119}+ x^{118}+ x^{117}+ x^{116}+ x^{115}+ x^{114}+ x^{113}+ x^{112}+ x^{111}+ x^{110}+ x^{109}+ x^{108}+ x^{107}+ x^{106}+ x^{105}+ x^{104}+ x^{103}+ x^{102}+ x^{101}+ x^{100}+ x^{99}+ x^{98}+ x^{97}+ x^{96}+ x^{95}+ x^{94}+ x^{93}+ x^{92}+ x^{91}+ x^{90}+ x^{89}+ x^{88}+ x^{87}+ x^{86}+ x^{85}+ x^{84}+ x^{83}+ x^{82}+ x^{81}+ x^{80}+ x^{79}+ x^{78}+ x^{77}+ x^{76}+ x^{75}+ x^{74}+ x^{73}+ x^{72}+ x^{71}+ x^{70}+ x^{69}+ x^{68}+ x^{67}+ x^{66}+ x^{65}+ x^{64}+ x^{63}+ x^{62}+ x^{61}+ x^{60}+ x^{59}+ x^{58}+ x^{57}+ x^{56}+ x^{55}+ x^{54}+ x^{53}+ x^{52}+ x^{51}+ x^{50}+ x^{49}+ x^{48}+ x^{47}+ x^{46}+ x^{45}+ x^{44}+ x^{43}+ x^{42}+ x^{41}+ x^{40}+ x^{39}+ x^{38}+ x^{37}+ x^{36}+ x^{35}+ x^{34}+ x^{33}+ x^{32}+ x^{31}+ x^{30}+ x^{29}+ x^{28}+ x^{27}+ x^{26}+ x^{25}+ x^{24}+ x^{23}+ x^{22}+ x^{21}+ x^{20}+ x^{19}+ x^{18}+ x^{17}+ x^{16}+ x^{15}+ x^{14}+ x^{13}+ x^{12}+ x^{11}+ x^{10}+ x^{9}+ x^{8}+ x^{7}+ x^{6}+ x^{5}+ x^{4}+ x^{3}+ x^{2}+ x+ 1$. |

**Table.9. Respective Polynomial of IGFP8 and OGFP8 of the Given 8 bit S-Box.**

**3.3 Algorithm to generate S-Box from Polynomials over Galois field GF($2^{15}$) or GF($2^{255}$).**

**START.**

**Step OA.** Choose 4 Galois field Polynomials over Galois field GF($2^{15}$) or 8 Galois field Polynomials over Galois field GF($2^{255}$).

**Step.01.** If Number of Terms in BCNs are Half of Number of total terms Then Step 02. Else Step 0A.

**Step.02.** Convert to decimal the 4 or 8 bit binary number generated by bits in same position of 4 BCNs for Galois field Polynomials over Galois field GF($2^{15}$) or 8 Galois field Polynomials over Galois field GF($2^{255}$).

**STOP.**

**Time Complexity of the given Algorithm.** O(n).

3. **4 and 8 bit S-Box Generation by respective BCNs over Non Binary Galois Field GF($16^{15}$) and Galois Field GF($256^{255}$) respectively.** The coefficients of each polynomial over Non Binary Galois Field GF($16^{15}$) forms a 4-bit S-Box.

The Coefficient of highest or lowest degree term must be the 1st element in 4-bit S-box, the value of other elements are the value of coefficients with immediate degree less than or greater than the previous one. Let The Polynomial be,

**BP(x)** = $0x^{15}+1x^{14}+2x^{13}+3x^{12}+4x^{11}+5x^{10}+6x^{9}+7x^{8}+8x^{7}+9x^{6}+10x^{5}+11x^{4}+12x^{3}+13x^{2}+14x+15$…………..(ii)

For the above Polynomial The Constituted 4-bit S-Box have been given in Table 10.

| Row | Column | 1 | 2 | 3 | 4 | 5 | 6 | 7 | 8 | 9 | A | B | C | D | E | F | G |
|---|---|---|---|---|---|---|---|---|---|---|---|---|---|---|---|---|---|
| 1 | **Index** | 0 | 1 | 2 | 3 | 4 | 5 | 6 | 7 | 8 | 9 | A | B | C | D | E | F |
| 2 | **S-Box** | 0 | 1 | 2 | 3 | 4 | 5 | 6 | 7 | 8 | 9 | A | B | C | D | E | F |

**Table.10. constituted 4-bit Crypto S-Box.**

The Polynomial with coefficients in reverse order,

**BP(x)** = $15x^{15}+14x^{14}+13x^{13}+12x^{12}+11x^{11}+10x^{10}+9x^{9}+8x^{8}+7x^{7}+6x^{6}+5x^{5}+4x^{4}+3x^{3}+2x^{2}+1x+0$…………….(iii)

For the above Polynomial The Constituted 4-bit S-Box have been given in Table 11.

| Row | Column | 1 | 2 | 3 | 4 | 5 | 6 | 7 | 8 | 9 | A | B | C | D | E | F | G |
|---|---|---|---|---|---|---|---|---|---|---|---|---|---|---|---|---|---|
| 1 | **Index** | 0 | 1 | 2 | 3 | 4 | 5 | 6 | 7 | 8 | 9 | A | B | C | D | E | F |
| 2 | **S-Box** | 15 | 14 | 13 | 12 | 11 | 10 | 9 | 8 | 7 | 6 | 5 | 4 | 3 | 2 | 1 | 0 |

**Table.11. constituted 4-bit Crypto S-Box.**

The coefficients of each polynomial over Non Binary Galois Field $GF(256^{255})$ forms a 8-bit S-Box. The Coefficient of highest or lowest degree term must be the 1st element in 4-bit S-box, the value of other elements are the value of coef-ficients with immediate degree less than or greater than the previous one. Let The Polynomial be, Let the Polynomial be given in Table.12.,

| **Polynomial BP(x) =** |
|---|
| $0.x^{255}+ 1.x^{254}+ 2.x^{253}+ 3.x^{252}+ 4.x^{251}+ 5.x^{250}+ 6.x^{249}+ 7.x^{248}+ 8.x^{247}+ 9.x^{246}+ 10.x^{245}+ 11.x^{244}+ 12.x^{243}+ 13.x^{242}+ 14.x^{241}+$<br>$15.x^{240}+16.x^{239}+ 17.x^{238}+ 18. x^{237}+ 19.x^{236}+ 20.x^{235}+ 21.x^{234}+ 22.x^{233}+ 23.x^{232}+ 24.x^{231}+ 25.x^{230}+ 26.x^{229}+ 27.x^{228}+ 28x^{227}+$<br>$29.x^{226}+ 30.x^{225}+ 31.x^{224}+32. X^{223}+ 33. x^{222}+ 34.x^{221}+ 35.x^{220}+ 36.x^{219}+ 37.x^{218}+ 38.x^{217}+39. x^{216}+40. x^{215}+ 41.x^{214}+ 42.x^{213}+$<br>$43.x^{212}+ 44.x^{211}+ 45.x^{210}+ 46.x^{209}+ 47.x^{208}+48.x^{207}+ 49.x^{206}+ 50.x^{205}+ 51.x^{204}+ 52.x^{203}+ 53.x^{202}+ 54.x^{201}+ 55.x^{200}+ 56.x^{199}+$<br>$57.x^{198}+ 58.x^{197}+ 59.x^{196}+ 60.x^{195}+ 61.x^{194}+ 62.x^{193}+ 63.x^{192}+ 64.x^{191}+ 65.x^{190}+ 66.x^{189}+ 67.x^{188}+68.x^{187}+ 69.x^{186}+ 70.x^{185}+$<br>$71.x^{184}+ 72.x^{183}+ 73.x^{182}+ 74.x^{181}+ 75.x^{180}+ 76.x^{179}+ 77.x^{178}+ 78.x^{177}+ 79.x^{176}+80.X^{175}+ 81.x^{174}+ 82.x^{173}+ 83.x^{172}+84. x^{171}+$<br>$85.x^{170}+ 86.x^{169}+ 87.x^{168}+88. x^{167}+ 89.x^{166}+ 90.x^{165}+ 91.x^{164}+ 92.x^{163}+ 93.x^{162}+ 94.x^{161}+ 95.x^{160}+96.x^{159}+ 97.x^{158}+ 98.x^{157}+$<br>$99.x^{156}+ 100.x^{155}+ 101.x^{154}+ 102.x^{153}+ 103.x^{152}+ 104.x^{151}+105.x^{150}+106.x^{149}+ 107.x^{148}+108. x^{147}+109.x^{146}+110. x^{145}+$<br>$111.x^{144}+112.X^{143}+113.x^{142}+114.x^{141}+115.x^{140}+116.x^{139}+117. x^{138}+118. x^{137}+119. x^{136}+120. x^{135}+121.x^{134}+122.x^{133}+123.x^{132}+$<br>$124.x^{131}+125.x^{130}+ 126.x^{129}+ 127.x^{128}+128.x^{127}+129. x^{126}+ 130.x^{125}+131.x^{124}+ 132.x^{123}+133.x^{122}+134.x^{121}+ 135.x^{120}+$<br>$136.x^{119}+ 137.x^{118}+ 138.x^{117}+ 139.x^{116}+ 140.x^{115}+ 141.x^{114}+ 142.x^{113}+ 143.x^{112}+144.x^{111}+ 145.x^{110}+ 146.x^{109}+ 147.x^{108}+$<br>$148.x^{107}+ 149.x^{106}+ 150.x^{105}+ 151.x^{104}+ 152.x^{103}+ 153.x^{102}+ 154.x^{101}+ 155.x^{100}+ 156.x^{99}+ 157.x^{98}+ 158.x^{97}+ 159.x^{96}+160.x^{95}+$<br>$161.x^{94}+ 162.x^{93}+ 163.x^{92}+ 164.x^{91}+ 165.x^{90}+ 166.x^{89}+ 167.x^{88}+ 168.x^{87}+ 169.x^{86}+ 170.x^{85}+ 171.x^{84}+ 172.x^{83}+ 173.x^{82}+$<br>$174.x^{81}+ 175.x^{80}+176.x^{79}+ 177.x^{78}+ 178.x^{77}+ 179.x^{76}+ 180.x^{75}+ 181.x^{74}+ 182.x^{73}+ 183.x^{72}+ 184.x^{71}+ 185.x^{70}+ 186.x^{69}+$<br>$187.x^{68}+ 188.x^{67}+ 189.x^{66}+ 190.x^{65}+ 191.x^{64}+192.x^{63}+193. x^{62}+ 194.x^{61}+ 195.x^{60}+ 196.x^{59}+ 197.x^{58}+ 198.x^{57}+ 199.x^{56}+$<br>$200.x^{55}+ 201.x^{54}+ 202.x^{53}+ 203.x^{52}+ 204.x^{51}+ 205.x^{50}+ 206.x^{49}+ 207.x^{48}+208.x^{47}+ 209.x^{46}+ 210.x^{45}+ 211.x^{44}+ 212.x^{43}+$<br>$213.x^{42}+ 214.x^{41}+ 215.x^{40}+ 216.x^{39}+ 217.x^{38}+ 218.x^{37}+219.x^{36}+ 220.x^{35}+ 221.x^{34}+ 222.x^{33}+ 223.x^{32}+224.x^{31}+ 225.x^{30}+$<br>$226.x^{29}+ 227.x^{28}+ 228.x^{27}+ 229.x^{26}+ 230.x^{25}+ 231.x^{24}+ 232.x^{23}+ 233.x^{22}+ 234.x^{21}+ 235.x^{20}+ 236.x^{19}+ 237.x^{18}+ 238.x^{17}+$<br>$239.x^{16}+240.x^{15}+ 241.x^{14}+242. x^{13}+ 243.x^{12}+ 244.x^{11}+ 245.x^{10}+ 246.x^{9}+ 247.x^{8}+ 248.x^{7}+ 249.x^{6}+ 250.x^{5}+ 251.x^{4}+ 252.x^{3}+$<br>$253.x^{2}+ 254x+ 255.$ |

**Table.12. Polynomial to Construct 8-bit Identity S-Box.**

For the above Polynomial The Constituted 8-bit S-Box have been given in Table 13.

| **Row** | **Column** | 1 | 2 | 3 | 4 | 5 | 6 | 7 | 8 | 9 | A | B | C | D | E | F | G |
|---|---|---|---|---|---|---|---|---|---|---|---|---|---|---|---|---|---|
| 1 | **Index** | 0 | 1 | 2 | 3 | 4 | 5 | 6 | 7 | 8 | 9 | A | B | C | D | E | F |
| 2 | | 0 | 1 | 2 | 3 | 4 | 5 | 6 | 7 | 8 | 9 | 10 | 11 | 12 | 13 | 14 | 15 |
| 3 | | 16 | 17 | 18 | 19 | 20 | 21 | 22 | 23 | 24 | 25 | 26 | 27 | 28 | 29 | 30 | 31 |
| 4 | | 32 | 33 | 34 | 35 | 36 | 37 | 38 | 39 | 40 | 41 | 42 | 43 | 44 | 45 | 46 | 47 |
| 5 | | 48 | 49 | 50 | 51 | 52 | 53 | 54 | 55 | 56 | 57 | 58 | 59 | 60 | 61 | 62 | 63. |
| 6 | | 64 | 65 | 66 | 67 | 68 | 69 | 70 | 71 | 72 | 73 | 74 | 75 | 76 | 77 | 78 | 79 |
| 7 | | 80 | 81 | 82 | 83 | 84 | 85 | 86 | 87 | 88 | 89 | 90 | 91 | 92 | 93 | 94 | 95 |
| 8 | **S-Box** | 96 | 97 | 98 | 99 | 100 | 101 | 102 | 103 | 104 | 105 | 106 | 107 | 108 | 109 | 110 | 111 |
| 9 | | 112 | 113 | 114 | 115 | 116 | 117 | 118 | 119 | 120 | 121 | 122 | 123 | 124 | 125 | 126 | 127 |
| 10 | | 128 | 129 | 130 | 131 | 132 | 133 | 134 | 135 | 136 | 137 | 138 | 139 | 140 | 141 | 142 | 143 |
| 11 | | 144 | 145 | 146 | 147 | 148 | 149 | 150 | 151 | 152 | 153 | 154 | 155 | 156 | 157 | 158 | 159 |
| 12 | | 160 | 161 | 162 | 163 | 164 | 165 | 166 | 167 | 168 | 169 | 170 | 171 | 172 | 173 | 174 | 175 |
| 13 | | 176 | 177 | 178 | 179 | 180 | 181 | 182 | 183 | 184 | 185 | 186 | 187 | 188 | 189 | 190 | 191 |
| 14 | | 192 | 193 | 194 | 195 | 196 | 197 | 198 | 199 | 200 | 201 | 202 | 203 | 204 | 205 | 206 | 207 |
| 15 | | 208 | 209 | 210 | 211 | 212 | 213 | 214 | 215 | 216 | 217 | 218 | 219 | 220 | 221 | 222 | 223 |
| 16 | | 224 | 225 | 226 | 227 | 228 | 229 | 230 | 231 | 232 | 233 | 234 | 235 | 236 | 237 | 238 | 239 |
| 17 | | 240 | 241 | 242 | 243 | 244 | 245 | 246 | 247 | 248 | 249 | 250 | 251 | 252 | 253 | 254 | 255 |

**Table.13. Constituted Identity 8-bit S-Box.**

**Note.** The 32-bit S-Boxes can be constituted by Polynomials over Galois field $GF[(2^{32})^{(2\wedge32-1)}]$ and the 64-bit S-Boxes can be constituted by Polynomials over Galois field $GF[(2^{64})^{(2\wedge64-1)}]$.

4. **Conclusion.** From this Research Article in can be concluded that 4, 8, 32, 64 bit Substitution boxes can be constituted using Basic Polynomials or BPs over Galois Fields $GF(16^{15})$, $GF(256^{255})$, $GF[(2^{32})^{(2\wedge32-1)}]$ and $GF[(2^{64})^{(2\wedge64-1)}]$ respectively. For this reason Generation of 4, 8, 32, 64 bit S-Boxes have been generated very easily with very less complexity. It is a very important work in modern cryptography since the work has been dedicated to crypto community for upgradation of complexity of crypto algorithms and ease to develop smart and intelligent devices for future.

5. **Acknowledgement.** For this work I would like to acknowledge my Supervisor Dr. Ranjan Ghsoh for his continuous encouragement and support. I would also like to acknowledge Prof. Debatosh Guha for providing me the infrastructure to do this work very elegantly.

## 6. References.